\documentclass[12pt,epsf]{article}
\usepackage{epsfig}

\renewcommand{\thanks}[1]{\footnote{#1}} 

\newcommand{\be}{\begin{equation}}
\newcommand{\ee}{\end{equation}}
\newcommand{\bea}{\begin{eqnarray}}
\newcommand{\eea}{\end{eqnarray}}

\begin{document}

\pagestyle{empty}

\bigskip\bigskip
\begin{center}
{\bf \large An Introduction of Multiple Scales in a
Dynamical Cosmology}
\end{center}

\begin{center}
James Lindesay\footnote{e-mail address, jlslac@slac.stanford.edu} \\
Computational Physics Laboratory \\
Howard University,
Washington, D.C. 20059 
\end{center}

\begin{center}
{\bf Abstract}
\end{center}
The discovery of scale acceleration evidenced from
supernovae luminosities and spatial flatness of
feature evolution in the cosmic microwave background
presents a challenge to the understanding of the evolution
of cosmological vacuum energy.  Although some
scenarios prefer a fixed cosmological constant with
dynamics governed in a Friedman-Robertson-Walker (FRW)
geometry, an early inflationary epoch remains a popular
model for cosmology.  It is therefore advantageous to
develop a metric framework that allows a transition
from an early inflationary period to a late stage
dominated by dark energy.  Such a metric is here
developed, and some properties of this metric are
explored.

\setcounter{equation}{0}
\section{Introduction}
\indent

A cosmological constant is the favored scenario for a quantum
cosmology with a single, late time deSitter geometry.  For
such models, any initial quantum state with an energy density
determined in the absence of microscopic particulate excitations
(ie the particle vacuum) undergoes a transition to thermal
particles, with the cosmological constant generated by the
vacuum modes of the initial collective state\cite{NSBP06}. 
The metric of such a system should be described by the
standard Friedman-Robertson-Walker (FRW) geometry,
with an additional true cosmological constant term in the
Einstein equation connecting the geometry to the energy
content.  However, if there is an early inflation stage,
the cosmology should involve a scale transition connecting
the early time inflation scale to a later deSitter scale.  In such
a scenario, a true cosmological constant cannot represent the
early evolution of the cosmology, and can at most represent
a late time behavior of the cosmology.

\setcounter{equation}{0}
\section{Fluid Cosmology}
\indent

It will be assumed that the dynamics of the cosmology can be 
accurately described by the Einstein equation during the period
under consideration:
\be
G_{\mu \nu} \equiv \mathbf{R}_{\mu \nu} -
{1 \over 2} g_{\mu \nu} \mathbf{R} = - \left (
{8 \pi G_N \over c^4} T_{\mu \nu} + \Lambda g_{\mu \nu}
\right ).
\label{EinsteinEqn}
\ee
For the present discussion, the
cosmology evolves in the absence of any
true cosmological constant $\Lambda_{true}=0$.
For an ideal fluid (no dissipation), the energy-momentum tensor
takes the form
\be
T_{\mu \nu} = P \: g_{\mu \nu} + (\rho + P) u_\mu u_\nu,
\label{Tmunu}
\ee
where the four velocity of the fluid satisfies the consistency
condition
\be
u_\mu g^{\mu \nu} u_\nu = -1.
\label{consistency}
\ee
We assume isotropic flow $u_\theta = 0 = u_\phi$. 
The trace of the energy-momentum tensor is given by
$g^{\mu \nu} T_{\mu \nu} \equiv T_\mu ^\mu = 3 P - \rho$.
This gives the form of the pressure  and density in terms of
geometric quantities:
\be
\begin{array}{l}
P=T_\phi ^\phi= T_\theta ^\theta = -{c^4 \over 8 \pi G_N} G_\theta ^\theta , \\
\rho=3 P + {c^4 \over 8 \pi G_N} G_\mu ^\mu .
\end{array}
\label{Einsteindensity}
\ee
Likewise, the time and radial components of the flow field can be determined
to satisfy
\be
\begin{array}{l}
u_0 ^2 = \left ( T_{00} - g_{00} P \right ) / (\rho + P) 
=- \left ( {c^4 \over 8 \pi G_N} G_{00} + g_{00} P \right ) / (\rho + P) \\
u_r ^2  = \left ( T_{rr} - g_{rr} P \right ) / (\rho + P) 
= - \left ( {c^4 \over 8 \pi G_N} G_{rr} + g_{rr} P \right ) / (\rho + P) .
\end{array}
\ee

For later comparison, for the Friedman-Robertson-Walker geometry
\be
g_{\mu \nu} = - c^2 dt^2 
+ R^2 (ct) \left ( {dr^2 \over 1 - \kappa r^2}+ 
r^2 d \theta ^2 + r^2 sin^2 \theta d\phi ^2   \right ) ,
\ee
these fluid parameters take the form
\be
\begin{array}{l}
\rho= {3 c^4 \over 8 \pi G_N} \left [
{\kappa \over R^2} + \left (
 {\dot{R} \over R }    \right )^2 \right ]  , \\ \\
P=  - {c^4 \over 8 \pi G_N} \left [
{\kappa \over R^2} +  \left (
 {\dot{R} \over R }  \right ) ^2  + 
2 {\ddot{R} \over R}\right ] , \\ \\
u_0 = -1 \quad , \quad u_r = 0 .
\end{array}
\ee

\setcounter{equation}{0}
\section{Multiple Cosmological Scales}

\subsection{The river model}
\indent

The so called ``river model" has been explored by
some authors\cite{rivermodel} to gain insight into
the nature of black holes.  In general, the metric
takes an off diagonal form
\be  
ds^2 = -dt_R ^2 + [dr - \beta (r) dt_R] ^2 +
r^2 (d \theta ^2 + sin^2 \theta d\phi ^2) .
\ee
The speed $\beta$ has been interpreted by
some to be the speed of radial outflow of the
space-time river through which objects move
using the rules of special relativity.
A transformation can be developed that diagonalizes
the metric.  Substitution of the form
\be
t_R = t_*  - \int ^ r {\beta (r') \over 1-\beta ^2 (r')}dr'
\ee
gives
\be  
ds^2 = -(1-\beta ^2 (r) )dt_* ^2 + {dr^2 \over 1 - \beta ^2 (r)} +
r^2 (d \theta ^2 + sin^2 \theta d\phi ^2) .
\ee
The river speed becomes luminal at the horizon associated
with the $(ct_* , r)$ coordinates.

Rather than exploring the horizon associated with a black
hole, the geometry of deSitter space is of interest for the
present discussion.    If one explores the geometry given
by the metric
\be
g_{\mu \nu} = - \left( 1 - { r^2 \over R_v ^2 (ct) }  \right )c^2 dt ^2 
-  {2 \,  r \over R_v  (ct) } c dt \, dr 
+  \left ( dr^2 + r^2 d \theta ^2 + r^2 sin^2 \theta d\phi ^2   \right ) ,
\label{dSmetric}
\ee
the fluid parameters generated in Eq. \ref{Einsteindensity}
are given by
\be
\begin{array}{c}
\rho= {3 c^4 \over 8 \pi G_N} \left (
{1 \over R_v}     \right )^2 , \\ \\
P + \rho =  - {c^4 \over 4 \pi G_N} {d \over dct} \left (
{1 \over R_v }   \right ) = 
{c^4 \over 4 \pi G_N} { \dot{R}_v \over R_v ^2}   , \\ \\
u_0 = -1 \quad , \quad u_r = 0 .
\end{array}
\label{dSpressure}
\ee
The fluid is seen to be at rest with respect to the space-time
coordinates, with a radial ``river" flow speed given by
$\beta = r / R_v (ct)$.

A static deSitter geometry can be further transformed into the
form of the Friedman-Robertson-Walker geometry by
comparing the metric forms
\be
\begin{array}{l}
ds^2 = -(1-r^2 / R_v ^2) c^2 dt^2 + 
{dr^2 \over 1-r^2 / R_v ^2} + 
r^2 (  d \theta ^2 +  sin^2 \theta d\phi ^2) \\ \\
ds^2 = - c^2 d\tilde{t}^2 + 
\Delta [ d\tilde{r}^2  + 
\tilde{r}^2 (  d \theta ^2 +  sin^2 \theta d\phi ^2)].
\end{array}
\label{deSitterFRW}
\ee
The radial scales are seen to satisfy
$r=\tilde{r} \Delta$ from angular constraints.
A coordinate tranformation can be found for
${\dot{\Delta} \over \Delta} = {1 \over R_v}$ and
$\dot{R}_v=0$, giving
\be
\begin{array}{l}
ct = c\tilde{t} - {R_v \over 2} log \left ( 1 - {\tilde{r} ^ 2  \over R_v^2} e^{2 c \tilde{t} /R_v} \right ) , \\
r= \tilde{r} e^{c \tilde{t} / R_v} ,
\end{array}
\ee
or alternatively
\be
\begin{array}{l}
c\tilde{t} = ct + {R_v \over 2} log \left ( 1 - {r ^ 2  \over R_v^2}  \right ) , \\ 
\tilde{r}= {r e^{-c t / R_v} \over \left( 1-{ r^2 \over R_v ^2} \right )^{1/2} }  .
\end{array}
\ee

\subsection{Dynamic horizon scale in an isotropic space}
\indent

Motivated by the river model, one can construct a metric
that incorporates the dynamic scales of FRW geometries
with asymptotic behaviors similar to deSitter geometries. 
Consider the following hybrid metric:
\be
\begin{array}{c}
g_{\mu \nu} = - \left( 1 - {R^2(ct) r^2 \over R_v ^2 (ct) }  \right )c^2 dt ^2 
- 2 {R^2(ct) r \over R_v  (ct) } c dt  \,  dr \\
+ R^2 (ct) \left ( dr^2 + r^2 d \theta ^2 + r^2 sin^2 \theta d\phi ^2   \right )  \\ 
= -  c^2 dt^2 + R^2(ct) \left( dr - {r \over R_v (ct)} c dt  \right )^2  \\
+ R^2 (ct) \left (  r^2 d \theta ^2 + r^2 sin^2 \theta d\phi ^2   \right ) .
\end{array}
\label{FRWdSmetric}
\ee
The radial null geodesics satisfy
\be
dr_L = \left ( \pm {1 \over R} + {r_L \over R_v} \right )  c dt,
\ee
with the particle horizon defined by the values of the scales
$R, R_v$ at $t=0$, with
$r_L (0) = 0$
The hydrodynamic parameters can be immediately calculated using
Eq. \ref{Einsteindensity}:
\be
\begin{array}{l}
\rho= {3 c^4 \over 8 \pi G_N} \left (
{1 \over R_v}  + {\dot{R} \over R }    \right )^2 , \\ \\
P = -\rho - {c^4 \over 4 \pi G_N} {d \over dct} \left (
{1 \over R_v} + {\dot{R} \over R }  \right ) , \\ \\
u_0 = -1 \quad , \quad u_r = 0 .
\end{array}
\label{FRWdensity}
\ee
In order to compactly represent the dynamics of these equations,
it is convenient to define the reduced cosmological scale
$\mathcal{R}$ using
\be
{ \dot{\mathcal{R}} \over \mathcal{R} } \equiv
{1 \over R_v} + { \dot {R} \over R}.
\ee
Combining terms in Eq. \ref{FRWdensity},
one obtains the first law of thermodynamics for
an adiabatic expansion in terms of this scale
\be
d (\rho \mathcal{R} ^3) = - P d \mathcal{R} ^3 .
\ee

The dynamics of the reduced cosmological scale is directly
determined by the energy content of the cosmology as
expressed in Eq. \ref{FRWdensity}.  During epochs dominated
by constant energy density, the dynamics is dominated by
static values for $R_v$, with $\dot{R} \approx 0$, and
$\mathcal{R} \simeq \mathcal{R}_v e^{ct/R_v}$.  If there is
an initial inflation followed by a long term adjustment towards
dark energy domination, one expects the micro-physics to modify
the dynamic content of the density in a manner that causes the
scales to behave as follows:
\be
\begin{array}{c l}
R_I \Leftarrow R_v  (ct) \Rightarrow R_\Lambda  & 
for \quad 0 \leftarrow t / \tau_I \rightarrow \infty \\
\rho_I \Leftarrow \rho (ct) = \rho_v + \rho_{thermal} \Rightarrow \rho_\Lambda  & 
for \quad 0 \leftarrow t / \tau_{I,c} \rightarrow \infty \\
0 \Leftarrow \dot{R}(ct) \Rightarrow 0  & 
for \quad 0 \leftarrow t / \tau_c \rightarrow \infty \\
R(0) \Leftarrow R(ct) \Rightarrow R_c  & 
for \quad 0 \leftarrow t / \tau_c \rightarrow \infty 
\end{array}
\ee
where the cosmological time scale is expected to be orders of
magnitude greater than the inflationary time scale $\tau_c >> \tau_I$.
It is non-trivial to develop a transformation that diagonalizes
the metric in Eq. \ref{FRWdSmetric} as was done for
the deSitter geometry in Eq. \ref{deSitterFRW}, 
unless the scales correspond in late times $R_c = R_\Lambda$.
During the intermediate epoch $\tau_I < t < \tau_c$, the dynamics
of the reduced scale $\mathcal{R}$ is essentially the same as
that of the scale $R$, driven by the thermal energy content.

The dynamics can be expressed solely in terms of the energy
content.  Using Eq. \ref{FRWdensity}
\be
{d \over d ct} \rho = -\sqrt{{24 \pi G_N \rho  \over c^4 }} (P + \rho) .
\label{rhodynamics}
\ee
The usual form for the equation of state of the thermal content will
be assumed:
\be
\begin{array}{l}
\rho=\rho_v + \rho_{thermal}  \\
P = P_v + P_{thermal} = -\rho_v + w \rho_{thermal} ,
\end{array}
\ee
where $w=1/3$ for radiation and $w=0$ for pressureless matter. 
Thus, Eq. \ref{rhodynamics} can be re-expressed
\be
{d \over d ct} \rho_{thermal} = -\sqrt{{24 \pi G_N  \over c^4 }
(\rho_v + \rho_{thermal} ) } (1+w) \rho_{thermal} .
\ee

Druing the intermediate (thermal) period $\tau_I << t << \tau_c $, one
assumes $\rho_v << \rho_{thermal}$ and 
$\left | {\dot{R} \over R } \right |  >> {1 \over R_v} \cong const$, the
thermal density satisfies
\be
\begin{array}{l}
{d \over d ct} \rho_{thermal} = -\sqrt{{24 \pi G_N   \over c^4 } } 
(1+w_*) \rho_{thermal} ^{3/2} \\ \\
\left(  {\rho_* \over \rho_{thermal}} \right ) ^{1/2} \cong
1 + \sqrt{{6 \pi G_N \rho_*  \over c^4 } } (1+w_*) (ct - ct_*) ,
\end{array}
\label{thermaldensity}
\ee
which are the same as the behaviors predicted by the
Friedman-Lemaitre equations during the
thermal period\cite{BigBang}.

During the very early $t<<\tau_I$ and very late $t>\tau_c$ epochs, the evolution
of the thermal component of density can likewise be determined:
\be
\begin{array}{l}
{d \over d ct} \rho_{thermal} \approx -\sqrt{{24 \pi G_N  \rho_v \over c^4 } } 
(1+w_*) \rho_{thermal}  \\ \\
log \left(  {\rho_{thermal} \over \rho_*} \right )  \approx
- {3 \over R_v} (1+w_*) (ct - ct_*) ,
\end{array}
\label{vacuumdensity}
\ee
clearly indicating exponential decrease in the thermal density during
these epochs.

As an aside, one can examine the growth of measurable dark energy
generated as vacuum energy during microscopic thermalization
of collective macroscopic motions.  If the dark energy is of the form
\be
\rho_{dark} \sim \sum_{modes} {1 \over 2} \hbar k v_p \sim
(k_{UV} ^ 4 - k_{IR}^4 (t)) \hbar v_p ,
\ee
the measurable infrared modes are expected to scale inversely
with the horizon, with essentially luminal speed.  Thus, the measurable dark
energy grows like 
\be
\begin{array}{l}
\dot{\rho}_{dark} \sim {\rho_{darkUV} \over t_{UV}} \left (
{t_{UV} \over t}\right ) ^4 \\
\rho_{dark} \sim \rho_\Lambda \left (
1 - \left ( {t_{UV} \over t}  \right ) ^3  \right ) .
\end{array}
\ee

\begin{center} \textbf{Curvature arguments} \end{center}
\indent

As a final question, the inclusion of curvature will be briefly
explored.  If curvature is introduced by the inclusion in the FRW
part of the metric Eq. \ref{FRWdSmetric} in the form
\be
\begin{array}{c}
g_{\mu \nu} = - \left( 1 - {R^2(ct) r^2 \over R_v ^2 (ct) }  \right )c^2 dt ^2 
- 2 {R^2(ct) r \over R_v  (ct) } c dt  \,  dr \\
+ R^2 (ct) \left ( { dr^2 \over 1-\kappa r^2} + r^2 d \theta ^2 + r^2 sin^2 \theta d\phi ^2   \right )  ,
\end{array}
\ee
an ``open" cosmology ($\kappa=+1$) is excluded by the consistency condition Eq. \ref{consistency}
with $(u_r)^2 < 0$,
whereas a ``closed" cosmology ($\kappa=-1$) must be finely tuned, as similarly can be
infered from the FRW cosmology\cite{JLHPNOct04}.  Therefore, fluid consistency constraints
exclude a nonvanishing value for $\kappa$ as a likely scenario.

\setcounter{equation}{0}
\section{Conclusions}
\indent

A form for a metric that incorporates evolution from an early
inflationary epoch through a thermal period towards a
final deSitter state has been developed.  The metric can be
used to explore the transition state associated with the
initial thermalization period with as close a correspondence
with the FRW geometry as desired by appropriate choice of
the relative cosmological scales.  If an initial inflation state
is confirmed, it is hoped that future work
utilizing this metric will give insight into the microscopic
contributions to the evolution of cosmological dark energy.

\begin{center}
\textbf{Acknowledgment}
\end{center}

The author would like to acknowledge J.D. Bjorken for
introducing him to the river model for black holes.

\end{document}